\documentclass[12pt]{article}
\input{amssymb.sty}

\def\thefootnote{\fnsymbol{footnote}}
\catcode`\@=11
\def\numberbysection{\@addtoreset{equation}{section}
        \def\theequation{\thesection.\arabic{equation}}}
\numberbysection

\def\beq{\begin{equation}}
\def\eeq{\end{equation}}
\def\bea{\begin{eqnarray}}
\def\eea{\end{eqnarray}}
\def\cqfd{\hskip 2truemm \vrule height2mm depth0mm width2mm}
\def\half{{1 \over 2}}
\def\e{\varepsilon}
\def\s{\sigma}
\def\z{\zeta}
\def\t{\theta}

\def\restnn#1{\langle #1 \rangle_{2n}}
\def\restn#1{\langle #1 \rangle_{n}}
\def\G{{\cal G}}
\def\L{{\Bbb L}}
\def\M{{\Bbb M}}
\def\N{{\Bbb N}}
\def\Q{{\Bbb Q}}
\def\Z{{\Bbb Z}}
\newtheorem{theo}{Theorem}

\newtheorem{lem}{Lemma}
\newtheorem{prop}{Proposition}
\begin{document}

\pagenumbering{alph}
\setcounter{page}{0}
\renewcommand{\thefootnote}{\alph{footnote}}
\renewcommand{\rho}{\varrho}

\rightline{UCL--IPT--98--13}

\vskip 3cm
{\LARGE \centerline{On Parity Functions in Conformal Field Theories}}

\vskip 2cm

\centerline{\large D. Altsch\"uler$^1$, P. Ruelle$^2$\footnote{\ \ Chercheur
Qualifi\'e FNRS} and E. Thiran$^2$}

\vskip 1.5truecm
\centerline{$^1$ D\'epartement de Math\'ematiques}
\centerline{Universit\'e de Nantes}
\centerline{BP 92208}
\centerline{F--44322 \hskip 0.5truecm Nantes Cedex 03, France}
\bigskip \bigskip
\centerline{$^2$ Institut de Physique Th\'eorique}
\centerline{Universit\'e Catholique de Louvain}
\centerline{B--1348 \hskip 0.5truecm Louvain-La-Neuve, Belgium}

\vskip 3truecm
\begin{abstract}
We examine general aspects of parity functions arising in rational conformal
field theories, as a result of Galois theoretic properties of modular
transformations. We focus more specifically on parity functions associated with
affine Lie algebras, for which we give two efficient formulas. We investigate
the consequences of these for the modular invariance problem.
\end{abstract}

\renewcommand{\thefootnote}{\arabic{footnote}}
\setcounter{footnote}{0}
\newpage
\pagenumbering{arabic}
\baselineskip=15pt

\section{Introduction and notations}

Modular invariance has become a major tool in the ambitious program of
classifying all rational conformal field theories (RCFT). At genus one,
it is the statement that a RCFT can be put on a torus in a consistent way, so
that e.g. the partition function should be well--defined over the conformal
classes of tori \cite{cardy}. Since the seminal ADE classification of the
Wess--Zumino--Novikov--Witten (WZNW) models based on $su(2)$
\cite{ciz}, there has been much progress on this question, especially during
the last few years, which have seen arithmetical techniques come into play. In
particular, the technical analysis of the conditions expressing the
modular invariance of the partition function on the torus has shown that the use
of Galois theory leads to powerful restrictions. These
restrictions are now usually referred to as parity selection rules. They have
had a crucial role in various classification results, that of the
$su(3)$--based WZNW being among the most convincing \cite{gannon}.

This paper is devoted to the study of general properties of the parity
selection rules corresponding to the best--known RCFTs, namely the
WZNW models. We will be slightly more general, and will consider theories with
symmetry algebras given by isomorphic chiral affine Lie algebras. We give
several formulas for the corresponding parity functions, and present some
consequences of them.

We first fix the notations regarding affine Lie algebras (referring to
\cite{kac} for further details) and recall their modular properties. We denote
by $\G$ a finite simple Lie algebra. The untwisted level $k$ affine algebra
$\widehat \G_k$ based on $\G$ is generated by $\G$--valued currents $J(z)$
satisfying the following commutation rules
\beq
\Big[\langle T^a,J(z) \rangle,\langle T^b,J(w) \rangle \Big] =
\Big\langle [T^a,T^b],J(z) \Big\rangle \,\delta(z-w) + k \langle T^a,T^b
\rangle
\, \partial_z \delta(z-w),
\eeq
where $\{T^a\}$ is a set of generators for $\G$. When $k \geq 0$ is an integer,
the algebra $\widehat \G_k$ has a finite number of unitary irreducible
representations $L(p)$, labelled by the strictly dominant weights of
$\G$ in the alc\^ove $P^n_{++}(\G)$
\beq
P^n_{++}(\G) = \Big\{ p = (a_1,a_2,\ldots) \;:\; a_i > 0, \; {\rm and} \;
\sum_i k_i^\vee a_i < n \Big\},
\eeq
where $k^\vee_i$ are the Kac labels given by the decomposition of the highest
root into simple roots $\psi = \sum_i k^\vee_i \alpha_i$, and where we have set
$n=k+h^\vee$ with $h^\vee = \rho \cdot \psi + 1$ the dual Coxeter number of
$\G$, and $\rho$ half the sum of the positive roots. The normalization of the
scalar product is such that $\psi^2=2$. In the sequel we will almost exclusively
use the integer $n$, called the height, instead of $k$. We let $\chi_p(\tau)$ be
the specialized character of $L(p)$.

The alc\^ove $P_{++}^n$ is an affine Weyl chamber, that is, it is the quotient
of the weight lattice of $\G$ minus the union of all affine walls by the action
of the affine Weyl group $\widehat W_n(\G)$ of height $n$. Since the affine Weyl
transformations $\hat w$ have a well--defined parity, one can associate to any
weight $p$ a number $\e_n(\G;p)$ as follows:
\beq
\e_n(\G;p) = \cases{ 0 & if $p$ is in an affine wall, \cr \noalign{\smallskip}
+1 & if $\hat w(p) \in P_{++}^n$ for an even $\hat w$, \cr \noalign{\smallskip}
-1 & if $\hat w(p) \in P_{++}^n$ for an odd $\hat w$. \cr}
\eeq
For obvious reasons, $\e_n(\G;p)$ will be called the affine parity of $p$
(relative to $\widehat W_n(\G)$). It is well--defined on the weight lattice on
account of the fact that $\widehat W_n(\G)$ fixes the set of affine walls, and
has a free action elsewhere. It satisfies the following properties:
\beq
\e_n(\G;\hat w(p)) = ({\rm det}\hat w) \, \e_n(\G;p), \qquad
\e_n(\G;p+n\alpha^\vee) = \e_n(\G;p) \hbox{  for any co--root $\alpha^\vee$}.
\eeq

The Hilbert space of a conformal theory with symmetry algebra $\widehat \G_k
\times \widehat \G_k$ consists of representations $L(p) \otimes L(p')$ taken
with certain multiplicities $N_{p,p'}$
\beq
{\cal H} = \bigoplus_{p,p'} \; N_{p,p'} \, (L(p) \otimes L(p')),
\qquad N_{p,p'} \in \N.
\eeq
When the theory is put on a torus of modulus $\tau$, the partition function
takes the form \cite{cardy}
\beq
Z(\tau,\tau^*) = \sum_{p,p'} \; N_{p,p'} \, \chi_p(\tau) \, \chi_{p'}^*(\tau).
\label{partition}
\eeq
Since two tori with moduli $\tau$ and $a\tau + b \over c\tau + d$ for $\left(
{a \atop c} \; {b \atop d} \right) \in PSL(2,\Z)$, are conformally equivalent,
a consistency condition is that the partition function must be modular
invariant,
that is, $Z(\tau) = Z({a\tau + b \over c\tau + d})$. The modular group
$PSL(2,\Z)$ being generated by $\tau \rightarrow \tau + 1$ and $\tau \rightarrow
{-1 \over \tau}$, it is sufficient to check the invariance of $Z(\tau)$ under
these two substitutions.

For affine Lie algebras, it is known that the characters carry a linear
representation of the modular group \cite{kac} (the same is true of
all known RCFTs, although no general proof exists). Explicitly, one has
\beq
\chi_p(\tau +1) = \sum_{p'} \;T_{p,p'} \, \chi_{p'}(\tau), \qquad
\chi_p({\textstyle{-1 \over \tau}}) = \sum_{p'} \;S_{p,p'} \, \chi_{p'}(\tau).
\eeq
with $T$ and $S$ both symmetric and unitary. $T$
is diagonal with roots of unity on the diagonal, while $S$ is more complicated.
The crucial property for what follows is that $S$, like $T$,
has all its entries in a cyclotomic extension of the rationals (if one assumes
the existence of unitary matrices $S$ and $T$, this is in fact true in any RCFT,
as proved in \cite{cg}). This implies that the algebraic extension $\M \equiv
\Q(S_{p,p'})$  generated by the coefficients of $S$ is a Galois extension with
Abelian Galois group. $\M$ contains the sub--field $\L \equiv
\Q(S_{p,p'}/S_{p,\rho})$, of which $\M$ is at most a quadratic extension (by
$S_{\rho,\rho}$). The action on $S$ of the Galois group of $\M$ is 
particularly simple. Take $\s \in {\rm Gal}(\M/\Q)$. It has been shown \cite{cg}
that $\s$ induces a permutation of the weights in $P^n_{++}$, such that
\beq
\s(S_{p,p'}) = \e_\s(p) S_{\s(p),p'} = \e_\s(p') S_{p,\s(p')},
\quad \e_\s(p) \in \{\pm 1\}.
\label{galois}
\eeq
Because $S_{p,p'}^2 \in \L$, the permutation of $P_{++}^n$ induced by $\s$ is
determined only through its restriction to Gal($\L/\Q$). The numbers $\e_\s(p)$,
called Galois parities, are not representations of the Galois group, but rather
cocycles, satisfying $\e_{\s\s'}(p) = \e_\s(\s'(p)) \, \e_{\s'}(p)$. They are 
the central objects of this paper. In a general RCFT, the relations
(\ref{galois}) are still valid if we take  $p$ and $p'$ as labels for the set
$\cal P$ of primary fields.

If one inserts the modular transformations of the characters in
the partition function (\ref{partition}), and requires its modular invariance,
one obtains the condition that the matrix $N$ must commute with $T$ and $S$.
Then by acting with an element $\s$ of the Galois group of $\M$ on the equation
$[N,S]=0$, one obtains the important result that
\beq
N_{\s(p),\s(p')} = \e_\s(p) \, \e_\s(p') \, N_{p,p'}.
\eeq
The parity selection rules now follow from the requirement that the
coefficients of $N$ must be positive integers
\beq
\e_\s(p) \, \e_\s(p') = -1 \hbox{  for some $\s$ in Gal($\M/\Q$)} \quad
\Longrightarrow
\quad  N_{p,p'}=0.
\label{selection}
\eeq
On the other hand, if $\e_\s(p) \, \e_\s(p') = +1$ for all $\s$, then
$N_{p,p'}$ can be non--zero, in which case we will say that there is a coupling
between $p$ and $p'$.

Therefore, in order to know which $N_{p,p'}$ can be non--zero and which are to
vanish, it is of paramount importance to solve the parity equation, {\it i.e.}
to know all pairs of weights $(p,p')$ that satisfy
\beq
\e_\s(p) = \e_\s(p'), \qquad \hbox{for all }\s.
\label{parity}
\eeq
This equation is really the key ingredient to all known classification
results, but (hence ?) is notoriously hard to solve. 

These selection rules hold in any RCFT in which the characters transform in a
unitary representation of the modular group. They put very strong restrictions
on the multiplicities of the representations (of whichever algebra is present)
that build the Hilbert space, thus on the field content of the theory. Note that
they have a purely group theoretical origin, as the parity functions are fixed
once the chiral algebras hence the characters are fixed. In case the left and
right chiral algebras are not isomorphic, restrictions like (\ref{selection})
apply, if appropriate parity functions are used. We end this introductory 
section by making these functions explicit for affine Lie algebras.

In the case of affine Lie algebras, it is known that $S$ is equal to \cite{kac}
\beq
S_{p,p'} = \gamma(\G,n) \sum_{w \in W(\G)} ({\rm det}w) \,
e^{-2i\pi p \cdot w(p')/n}.
\eeq
with $W$ the finite Weyl group, and $\gamma(\G,n)$ a numerical constant. The
numbers $S_{p,p'}$ belong to the cyclotomic extension $\Q(\zeta_{nQ})$
---$\zeta_m$ will denote a primitive $m$\/-th root of unity---, for some
integer $Q$ depending on $\G$ (and possibly on $n$, see \cite{bcldb,ganwal}).
The elements of Gal$(\M/\Q)$ are indexed by  integers $h$ coprime with $nQ$,
{\it i.e.} by elements of $\Z^*_{nQ}$. The Euler function $\varphi(nQ)$ gives
the order of
$\Z^*_{nQ}$.

{}From the formula for $S_{p,p'}$, it is not difficult to compute the
permutation of the alc\^ove induced by $\s_h$: $\s_h(p)$ is the only weight in
the alc\^ove whose image by an affine Weyl transformation is the dilated weight
$hp$ (multiplication componentwise). In other words, there exists a unique
$w_{h,p} \in W(\G)$ and a unique co--root $\alpha^\vee_{h,p}$ of $\G$ such that
$\s_h(p)  = w_{h,p}(hp) + n \alpha^\vee_{h,p}$. Moreover the Galois parity takes
the value
\beq
\e_{\s_h}(p) = {\textstyle {\s_h(\gamma(\G,n)) \over \gamma(\G,n)}}\e_n(\G;hp),
\eeq
which is an affine parity up a constant prefactor (itself a sign because
$[\gamma(\G,n)]^2 \in \Q$). Since this prefactor does not depend on $p$, it
clearly drops out of the selection rules (\ref{selection}) ---it would however
matter if the chiral algebras were not isomorphic---, so we neglect it from now
on (except in Section 3). Therefore the parity equation for affine Lie algebras
takes the form
\beq
\e_n(\G;hp) = \e_n(\G;hp'), \qquad \forall h \in \Z^*_{nQ}.
\label{affpar}
\eeq
Note that the map $h \longmapsto {\s_h(\gamma(\G,n)) \over \gamma(\G,n)} =\pm 1$
is a homomorphism, so that the affine parity $\e_n(\G;hp)$ itself is a cocycle.

An algorithm to compute the parity of an arbitrary weight can be given, that
requires evaluating congruences on Dynkin labels and determinants of
permutations (see \cite{rtw} for $\G=A_\ell$). It is not our purpose to
describe that algorithm in the general case since, as we shall soon see, $\G$
parities can be reduced to the much simpler $su(2)$ parities, which we now make
explicit.

In the Dynkin basis, an $su(2)$ weight is just an integer, and the weight
lattice is $\Z$. The dual Coxeter number is $h^\vee=2$ so that the alc\^ove at
height $n$ is the set
\beq
P_{++}^n(su(2)) = \{ a \in \Z \;:\; 1 \leq a \leq n-1 \}.
\eeq
The affine walls are the points of the ideal $n\Z$. The co--roots
correspond to even integers, which implies that the parity function of $su(2)$
is periodic with period $2n$. Therefore it only depends on the residue modulo
$2n$ of its argument, which we denote by $\restnn a$, taken between 0 and
$2n-1$. (More generally, we denote by $\langle x \rangle_y$ the residue of
$x$ modulo $y$, chosen in $[0,y-1]$.) Putting all together, we find for any
integer $a$
\beq
\e_n(a) \equiv \e_n(su(2);a) = \cases{ 0 & if $a=0 \bmod n$, \cr
+1 & if $\restnn {a} < n$, \cr
-1 & if $\restnn {a} > n$. \cr}
\eeq
This is confirmed by computing directly the action of the Galois group on the 
$S$ matrix, given for $su(2)$ by $S_{a,a'} = \sqrt{2 \over n} \sin{\pi aa' \over
n}$. For later use, we collect the main properties of the $su(2)$
parity:
\bea
&& \e_n(a) = {\rm sgn}\big(\sin{\pi a \over n}\big) =
2 - {\restnn {a} + \restnn {n-a} \over n}, \qquad a \not\in n\Z, \label{sine}\\
&& \e_n(a) = \e_n(n-a) = \e_n(a+2n) = -\e_n(-a). \label{auto}
\eea

To summarize, the main conclusion, as far as affine Lie algebras are concerned, 
is that the Galois parities coincide with the affine parities. Solving the 
parity equation (\ref{affpar}) is nonetheless extremely hard, which explains
why the general solution is known for $su(2)$\footnote{At the time the
classification  of affine $su(2)$ modular invariant partition functions was
completed \cite{ciz}, the Galois symmetry of the $S$ matrix had not yet been
recognized, and consequently there was no parity equation. The now available
general solution of the $su(2)$ parity equation would yield the result in a
more efficient way.} and $su(3)$ only. For $su(2)$, the result is fairly simple,
even though the proof is not completely straightforward, despite the deceptive
simplicity of the parity function. In the case of $su(3)$, the parity equation
is considerably more complex, and it is only recently that the general solution
has been given \cite{aoki}, though in a totally different context. As noticed 
in \cite{rtw}, the $su(3)$ parity plays a fundamental role in
the description of the Jacobian varieties of the complex Fermat curves, and it
is in this geometric setting that, in disguise, the equation for $su(3)$ was
solved in all generality (see \cite{bcir} for a review of the connections
between the two problems). The $su(3)$ solution yields, as a special case, the
solution for the $su(2)$ case. For higher rank algebras, virtually nothing is
known about the parity equation. 

It is our purpose here to suggest new directions, by showing that some of the
properties that proved very useful for the $su(2)$ and $su(3)$ algebras, in
fact go over to the other cases.

One may also note that focussing on $su(2)$ parities is not only important for
dealing with parities arising in affine algebras. They turn out to be relevant
in other models as well. Good examples are
provided by minimal conformal theories ${\cal M}(p,q)$, in which the Galois
parities are just products of two $su(2)$ parities, taken at heights $p$ and
$q$. Because the $S$ matrices in rational conformal theories are often related
to sine functions, $su(2)$ parities inevitably emerge when acting with the
Galois groups. This should be no surprise as most rational theories can be
constructed as cosets of WZWN models.


\section{Formulas for parities}

We will present in this section two explicit formulas to compute the parity
functions in affine algebras. They have very different flavours, one being
multiplicative, the other additive. Perspectives offered by these formulas
are investigated in the subsequent sections.

The first, multiplicative, formula relates the parity in any (untwisted) affine
algebra to the parity function in the simplest of all, namely $su(2)$. For $p$
a weight of $\G$, not necessarily dominant, the following formula yields an
expression for the parity of $p$ relative to the affine Weyl group $\widehat
W_n(\G)$
\beq
\e_n(\G;p) = \prod_{{\rm roots\ }\alpha >0}\,\e_{nD}(su(2);D\alpha \cdot p) =
\prod_{\alpha >0}\,{\rm sgn}\left(\sin{\pi \alpha \cdot p \over n}\right),
\label{product}
\eeq
where $D$ is the smallest positive integer such that $D \alpha \cdot p \in \Z$
for all weights $p$ and all roots $\alpha$. Explicitly $D=1$ for $\G$
simply--laced, $D=2$ for $\G=B_\ell,C_\ell, F_4$, and $D=3$ for $\G=G_2$.

The proof of the product formula (\ref{product}) is not difficult. One may first
check that both expressions coincide when $p$ is in the fundamental alc\^ove
$P_{++}^k(\G)$ (clear because $p$ in the alc\^ove implies $\alpha \cdot p \in
[1,n-1]$), and then verify that they have the same transformation properties
under the affine Weyl group. For the translational part, one uses, for any
co--root $\alpha^\vee$,
\beq
{\e_n(\G;p+n\alpha^\vee) \over \e_n(\G;p)} =
\prod_{\alpha >0} {\e_{nD}\big(D\alpha \cdot p + nD\alpha \cdot
\alpha^\vee \big) \over \e_{nD}(D\alpha \cdot p)} = \prod_{\alpha >0}
(-1)^{\alpha \cdot \alpha^\vee} = (-1)^{2\rho \cdot \alpha^\vee} = +1.
\eeq
For the finite Weyl part, one checks
\beq
\prod_{\alpha >0} \e_{nD}(D\alpha \cdot w(p)) =
\prod_{\alpha >0} \e_{nD}(Dw^{-1}(\alpha) \cdot p) =
(-1)^{t_w} \prod_{\alpha >0} \e_{nD}(D\alpha \cdot p) =
({\rm det}w) \prod_{\alpha >0} \e_{nD}(D\alpha \cdot p),
\eeq
with $t_w$ the number of positive roots whose image under $w$ are negative
roots.

Alternatively one may obtain the formula (\ref{product}) by acting with an
element of the Galois group Gal$(\M/\Q)$ on the factorized form for the $S$
matrix elements
\beq
S_{\rho,p}(\G) = \gamma'(\G) \prod_{\alpha >0} S_{\rho,\alpha \cdot p}(su(2)),
\eeq
for some constant $\gamma'(\G)$ that only depends on $\G$.

Our second formula is additive and has a stronger arithmetical taste. According
to the previous, multiplicative expression, parity functions in affine algebras
are products of $su(2)$ parities $\e_n(\alpha \cdot p)$ (say when $D=1$). As
mentioned before, these $su(2)$ parities depend on the residues of their
argument modulo $2n$. However, in the particular case $\G=su(3)$, the
parity function, a product of three $su(2)$ parities according to
(\ref{product}), 
\beq
\e_n(su(3);p) = \e_n(a)\e_n(b)\e_n(a+b) = \e_n(a)\e_n(b)\e_n(n-a-b),
\qquad p=(a,b)
\eeq
can also be written in a way that only involves residues modulo $n$. Indeed one
may check that
\beq
\e_n(su(3);p) = \left\{+1 \atop -1 \right\}
\qquad \Longleftrightarrow \qquad \restn{a} + \restn{b} + \restn{n-a-b} =
\left\{n \atop 2n \right\}.
\label{a2add}
\eeq

Since this  additive formula proved extremely useful to solve the parity
equation for $su(3)$ \cite{kr,aoki}, it is a natural question to see
if it can be generalized. It can indeed be generalized, though not uniformly
for all algebras, the resulting formulas being dependent of the structure of
the root systems. They are primarily based on the following basic observation.
\begin{lem}
Suppose that $x_1,x_2,\ldots,x_m$ are integers in $\Z \,\backslash\,
n\Z$ satisfying $\sum_i x_i = \delta n \bmod 2n$, with $\delta=0,1$. Then
\beq
\e_n(x_1) \e_n(x_2) \ldots \e_n(x_m) = (-1)^\delta \left\{+1 \atop -1 \right\}
\qquad
\hbox{iff} \qquad \sum_i \;\restn{x_i} = \left\{0 \atop n \right\} \bmod 2n.
\eeq
\end{lem}

\noindent
{\sl Proof.} Let $\mu$ be the number of indices $i$ such that
$\e_n(x_i)=-1$. Since for those $i$'s, $\restn{x_i}=\restnn{x_i}-n$, we get the
following equalities modulo $2n$:
\beq
\sum_i \;\restn{x_i} = \sum_i \;\restnn{x_i} - \mu n =  (\delta+\mu)n \bmod 2n.
\eeq
On the other hand, $\prod_i \e_n(x_i) = (-1)^\mu$, which proves the lemma.
\cqfd

\medskip
This simple result is the key to the generalization of (\ref{a2add}).
Let us first consider the algebras $su(N)$, for $N$ odd. Recall that a positive
root $\alpha$ of $su(N)$ has level $|\alpha|=l$ if $\alpha$ is the sum of $l$
simple roots, and that the set of positive roots of has the property that
$\sum_{|\alpha|=l} \alpha = \sum_{|\alpha|=N-l} \alpha$.  

For a weight $p=(a_1,a_2,\ldots,a_{N-1})$, the product formula (\ref{product})
says that the affine parity of $p$ is the product of $su(2)$ parities
$\e_n(\alpha \cdot p)$ over all positive roots. One can then satisfy the
hypothesis of Lemma 1 by replacing $\e_n(p \cdot \alpha)$ by $\e_n(n-p
\cdot \alpha)$ for all positive roots of level bigger or equal to
$N+1 \over 2$. Doing so, we obtain
\beq
\e_n(su(N);p) =
\prod_{\alpha >0 \atop |\alpha| \leq (N-1)/2} \e_n(p \cdot \alpha)
\prod_{\alpha >0 \atop |\alpha| \geq (N+1)/2} \e_n(n - p \cdot \alpha),
\qquad \hbox{$N$ odd}.
\eeq
The relevant value of $\delta$ is given by the number of positive roots whose
level is bigger or equal to $N+1 \over 2$, namely $\delta={N^2-1 \over 8} \bmod
2$. Thus the lemma yields the following.
\begin{prop}
For $N \geq 3$ odd, one has
\bea
&& \qquad \e_n(su(N);p) = (-1)^{(N^2-1)/8} \left\{+1 \atop -1
\right\} \qquad \hbox{iff} \nonumber\\
&& \hskip 3truecm \sum_{\alpha >0 \atop |\alpha| \leq (N-1)/2} \restn{p \cdot
\alpha} + \sum_{\alpha >0 \atop |\alpha| \geq (N+1)/2} \restn{n - p \cdot
\alpha}  = \left\{ 0 \atop n \right\} \bmod 2n. \hskip 1truecm
\label{Nodd}
\eea
\end{prop}

\medskip
For $N=3$, it reproduces (\ref{a2add}) because the sum $\restn{p \cdot
\alpha_1} + \restn{p \cdot \alpha_2} + \restn{n - p \cdot (\alpha_1+\alpha_2)}$
can take only two values, $n$ or $2n$. 

The same trick does not always work for other algebras, because
it relies on the fact that the positive roots can be partitioned into two
sets such that the sum of the roots in one set equals the sum of the
roots in the other set. In fact, it is not so much the roots which matter,
but their scalar products with $p$. So the condition underlying the above
proposition is the existence of two disjoints sets $A$ and $B$ such that
$\sum_{\alpha \in A} \alpha \cdot p = \sum_{\alpha \in B} \alpha \cdot p$. When
this is not possible, there are two alternatives. Either one constrains the
weight $p$ so that it be possible, or one takes suitable multiples of the 
height $n$. We illustrate it in $su(4)$, which is the simplest case for which
this occurs.

For $p=(a,b,c)$ a general weight of $su(4)$, the product formula yields 
\beq
\e_n(su(4);p) = \e_n(a) \e_n(b) \e_n(c) \e_n(a+b) \e_n(b+c) \e_n(a+b+c).
\label{a3parity}
\eeq
One checks that if $p$ is generic, there is no way to change some of the
arguments as above, in such a way that they sum up to a multiple of $n$. It is
however possible if $p$ is self--conjugate, $a=c$, since by inserting
$\e_n^2(a)=1$, one has
\beq
\e_n(su(4);p) = \e_n(b) \e_n(2a+b) \e_n^2(a) = \e_n(a) \e_n(a) \e_n(b)
\e_n(n-2a-b).
\eeq
A simple application of the lemma implies, for a self--conjugate weight
$p=(a,b,a)$, that 
\beq
\e_n(su(4);p) = +1 \qquad \hbox{\it iff} \qquad 2\restn{a} + \restn{b} +
\restn{n-2a-b} = n \bmod 2n.
\label{su4}
\eeq

If one wishes to keep a generic weight, the other way to proceed is to use the
obvious identity $\e_n(x)=\e_{2n}(2x)$, and then to insert $\e_{2n}^2(a)
\e_{2n}^2(c)=1$ in (\ref{a3parity}):
\bea
\e_n(su(4);p) \!\!\!&=&\!\!\! \e_{2n}(2a) \e_{2n}(2b) \e_{2n}(2c) \e_{2n}(2a+2b)
\e_{2n}(2b+2c) \e_{2n}(2a+2b+2c) \e_{2n}^2(a) \e_{2n}^2(c) \nonumber\\
\!\!\!&=&\!\!\! \e_{2n}(2a) \e_{2n}(2b) \e_{2n}(2c) \e_{2n}(2a+2b) \e_{2n}(c)
\e_{2n}(c) \times \nonumber\\
&& \quad \e_{2n}(2n-2b-2c) \e_{2n}(2n-2a-2b-2c) \e_{2n}(2n-a)
\e_{2n}(2n-a).
\eea
The lemma can be used once more to relate the affine parity of a general $su(4)$
weight to a sum of residues modulo $2n$. The price to pay is the larger number
of residues that now enter the formulas.

For the other $su(N)$ algebras, $N$ even, the first alternative
(self--conjugate weights) works if $N=0 \bmod 4$, while the second works well
for all $N$ even. Similar formulas can be designed for all other simple Lie
algebras.

In the following sections, we present some implications of the above
multiplicative and additive formulas.

In Section 3, we show that they allow various cohomological interpretations,
and implies certain relations between the field extensions $\M$ and $\L$. In
particular, as a sort of generalization of (\ref{product}), we prove a formula
expressing the affine parities for $su(2N+1)$ as products of $su(2N)$ parities,
which has a strong cohomological flavour. 

In terms of computational efficiency, the formula (\ref{product}) is much easier
to handle than the previously known formula, which requires computing the
parity of a Weyl transformation \cite{rtw}. As we shall see in Section 4, it
also clearly shows why certain non--trivial couplings are allowed by the parity
selection rules, and how conversely, trivial solutions to the parity equation
can give rise to non--trivial couplings, which could be otherwise hard to guess.
Moreover, we relate the solutions of the parity equation to the
existence of certain totally positive numbers in the field $\Q(\sin{\pi \over
n})$. This allows the construction of solutions which, we will argue, appear to
be the generic solutions. 

Finally in Section 5, we show that the additive formulas might reveal a new path
into solving the parity equation. At present, this last approach appears more
promising to us, in spite of the fact that difficult and deep arithmetical
questions seem to emerge on the way.


\section{Cohomological interpretations}

In this section we give cohomological interpretations of the relations
satisfied by the parities:
\beq
\e_{\sigma\sigma'}(j)=
\e_\sigma(j)\e_{\sigma'}(\sigma(j))=
\e_{\sigma'}(j)\e_{\sigma}(\sigma'(j)),
\label{cocycle}
\eeq
where $\sigma,\sigma' \in {\rm Gal}(\M/\Q)$, and $j$ labels the elements
of $\cal P$, the finite set of chiral primary fields. The second equality
follows from the fact that ${\rm Gal}(\M/\Q)$ is Abelian.
We begin by reviewing some definitions of group cohomology, for which we adopt
a multiplicative notation.

Let $G$ be a group, and $A$ be a multiplicative Abelian group.
Assume that $G$ acts on $A$ by automorphisms, {\it i.e.} there is
a homomorphism $\alpha : G \rightarrow {\rm Aut}(A)$.
For simplicity we write $g \cdot a$ instead of $\alpha(g)(a)$,
where $g\in G$, $a\in A$. The set $C^n(G,A)$ of $n$--cochains is the Abelian
group of functions which depend on $n$ variables in $G$ and with values in $A$:
\beq
C^n(G,A) = \Big\{\,f \;:\; \underbrace{G \times\cdots\times G}_{n \;{\rm
factors}} \rightarrow A \,\Big\}.
\eeq
By definition, a 0--cochain is a fixed element of $A$, so that $C^0(G,A)=A$.
One also defines coboundary operators $\delta_n : C^n \rightarrow C^{n+1}$,
which, for $n=0,1$, are given explicitly by
\bea
&& \Big(\delta_0(a)\Big)(g) = (g \cdot a)a^{-1}, \qquad g\in G, \;\; a\in A, \\
&& \Big(\delta_1(f)\Big)(g,h) = \Big(g \cdot f(h)\Big) f(g) f(gh)^{-1},
\qquad f\in C^1(G,A), \;\; g,h\in G.
\eea
The group of 1--coboundaries is $B^1(G,A)={\rm Im}(\delta_0)$, whereas the group
of 1--cocycles is $Z^1(G,A)= \ker(\delta_1)$. It is easy to see that $\delta_1
\circ \delta_0 = 1$, so that $B^1(G,A) \subset Z^1(G,A)$. The first cohomology
group is then $H^1(G,A)=Z^1(G,A)/B^1(G,A)$.

Now consider a RCFT with the finite set ${\cal P}$ of primary fields. 
Take $A=\{+1,-1\}^{\cal P}$ to be the multiplicative Abelian group of functions:
${\cal P} \rightarrow \{+1,-1\}$ (multiplication componentwise), and take 
$G={\rm Gal}(\M/\Q)$. As recalled in the Introduction, $G$ acts on ${\cal P}$ by
permutations $j \mapsto \s(j)$, and thus also on $A$ by $(\s \cdot
a)(j)=a(\s(j))$. The first equality in (\ref{cocycle}) then translates into the
property that the map $\e : G \rightarrow A$ defined by
$\sigma \mapsto \e_\sigma(\cdot)$ is a 1--cocycle in $C^1(G,A)$.

\begin{prop}
If $\e$ is a coboundary, $\M=\L$.
\end{prop}
{\em Proof.}
We know that ${\rm Gal}(\M/\L)$ is the kernel of the restriction ${\rm
Gal}(\M/\Q) \rightarrow {\rm Gal}(\L/\Q)$, therefore if $\sigma\in{\rm Gal}
(\M/\L)$, $\sigma(S_{ij})=\e_\sigma(i) S_{ij}$, since the permutation of $\cal
P$ induced by $\s$ is determined by its restriction to Gal($\L/\Q$). By the
assumption on $\e$, $\e_\sigma(i)=a(\sigma(i))/a(i)$, for some $a\in
A$, thus $\e_\sigma(i)=1$ if $\sigma\in{\rm Gal}(\M/\L)$.
Hence $\sigma(S_{ij})=S_{ij}$ for all $\sigma\in{\rm Gal}(\M/\L)$.
\cqfd

Examples of RCFTs where $\e$ is a coboundary include all models with the
current algebra $su(N^2)$ at level 1. For these cases one easily checks that
$\e_\s(p)=+1$ for all $\s$ and all $p$ in the alc\^ove, and indeed
$S_{\rho,\rho}={1 \over N}$ implies $\M=\L(S_{\rho,\rho})=\L$. (Note that
$\e_\s(\cdot)$ is the full parity defined in (\ref{galois}), and not the
affine parity $\e_n({\cal G};\cdot)$.) The converse is however not true: in
models with current algebra $su(2)$ at even level, it is known that $\M=\L$ (see
f.i. \cite{bcldb}) but $\e$ is never a coboundary\footnote{The field
extensions $\M$ and $\L$ have been determined in \cite{ganwal} for the current
algebras based on $su(N)$. Many of them have $\L=\M$.}.

For $j\in{\cal P}$, we denote by $G_j=\{\sigma\in G \;|\; \sigma(j)=j\}$ the
stabilizer of $j$. Note that since $G$ is Abelian, $G_j=G_k$ if $j$ and $k$
belong to the same orbit ${\cal O}$ of $G$ in ${\cal P}$, thus it
makes sense to define the stabilizer of an orbit ${\cal O}$ by $G_{\cal O}=G_j$
with $j\in {\cal O}$. Let $\widehat{G}_{\cal O}$ be the group of homomorphisms
$G_{\cal O}\rightarrow\{+1,-1\}$.
\begin{prop}
There is an embedding $H^1(G,A)\hookrightarrow \prod_{\cal O} \widehat{G}_{\cal
O}$, where the product is over all the orbits ${\cal O}$.
\label{p2}
\end{prop}

\noindent
The proof of proposition \ref{p2} is based on the following lemma:
\begin{lem}
$\e$ is a coboundary if and only if for all $j\in{\cal P}$
and all $\sigma\in G_j$, $\e_\sigma(j)=1$.
\end{lem}
{\em Proof.}
If we assume that $\e$ is a coboundary, then it is obvious that $\e_\sigma(j)=1$
if $\sigma(j)=j$. Assume now that $\e_\sigma(j)=1$ for all $\sigma\in G_j$.
We have to construct a function $a(j)$ such that $\e_\sigma(j) =
a(\sigma(j))/a(j)$.

First we observe that the cocycle condition (\ref{cocycle}) implies that
$\e_{\sigma\sigma'}(j)=\e_\sigma(j)$ if $\sigma'\in G_j$. Thus if we restrict 
$j$ to lie in a certain orbit ${\cal O}$, $\e_\sigma(j)$ depends only
on $\sigma\bmod G_{\cal O}$, and we can think of $\sigma$ as lying in 
$G/G_{\cal O}$.

Let us choose one particular element $j_0$ as the origin of ${\cal O}$. Every
$j\in{\cal O}$ can be written in a unique way as $j=\sigma\cdot j_0$ for some
$\sigma\in G/G_{\cal O}$. We define the restriction of $a$ to ${\cal O}$ by
$a(j)=\e_\sigma(j_0)$. From (\ref{cocycle}) we get
\beq
\e_{\sigma\sigma'}(j_0) = \e_{\sigma'}(j_0)\e_\sigma(\sigma'\cdot j_0),
\eeq
so that upon setting $k=\sigma'\cdot j_0$, we get
\beq
\e_\sigma(k)=\e_{\sigma\sigma'}(j_0)/\e_{\sigma'}(j_0) =
a(\sigma(k))/a(k). \cqfd
\eeq

\noindent
{\em Proof of proposition \ref{p2}.}
We consider the second equality in (\ref{cocycle}), and assuming 
that $\sigma\in G_j$, we obtain $\e_\sigma(j)=\e_\sigma(\sigma'\cdot j)$.
Therefore if $\sigma\in G_{\cal O}$, $\e_\sigma(\cdot)$
is constant on ${\cal O}$. Denote this constant by $\e_\sigma({\cal O})$. It is
easy to see from (\ref{cocycle}) again, that $\sigma\mapsto\e_\sigma({\cal O})$
belongs to $\widehat{G}_{\cal O}$. Thus we have now a map
\beq
\tilde{\rho} : Z^1(G,A) \rightarrow \prod_{\cal O} \widehat{G}_{\cal O}.
\eeq
The easy direction of the lemma says that $B^1(G,A)\subset\ker(\tilde{\rho})$,
so that $\tilde{\rho}$ descends to a map
\beq
\rho : H^1(G,A) \rightarrow \prod_{\cal O} \widehat{G}_{\cal O},
\eeq
and the other direction says that in fact
$B^1(G,A)=\ker(\tilde{\rho})$, so that $\rho$ is injective. \cqfd

\medskip
We close this section by mentionning another product formula, relating the
affine parities of $su(2N)$ and $su(2N+1)$. Formally, the formula says that, in
the appropriate cohomology, the affine parity of $su(2N+1)$ is the coboundary
of the affine parity of $su(2N)$, both algebras taken at the same height:
\bea
&& \hskip -1truecm \e_n\Big(su(2N+1);(a_1,a_2,\ldots,a_{2N})\Big) = 
\hbox{`` }\delta_{2N-1}\,\e_n\Big(su(2N);\cdot\Big) \hbox{ ''} \\
&& = \e_n\Big(su(2N);(a_2,\ldots,a_{2N})\Big) \prod_{i=1}^{2N-1}
\e_n\Big(su(2N);(a_1,\ldots,a_i+a_{i+1},\ldots,a_{2N})\Big) \nonumber\\
&& \qquad \qquad \times \; \e_n\Big(su(2N);(a_1,\ldots,a_{2N-1})\Big).
\eea
It is only a formal coboundary since, on $\Z^{2N-1}$, the parity
$\e_n(su(2N);\cdot)$ takes the values $\{0,+1,-1\}$, which is not a
multiplicative group. Nevertheless, in terms of affine parities, it yields an
identity whose proof is straightforward: the two expressions are equal to
$+1$ when $p=(a_1,a_2,\ldots,a_{2N})$ is in the alc\^ove $P_{++}^n(su(2N+1))$, 
and they transform the same way under the affine Weyl group $\widehat
W_n(su(2N+1))$. At this level of generality, these identities seem to be
specific to the $A_l$ series, even if other relations can be found. For
instance, the $su(5)$ parity for a general weight is the product of four $su(3)$
parities, while a $G_2$ parity is the product of two $su(3)$ parities.


\section{Totally positive numbers}

For affine Lie algebras, the parity equation (\ref{affpar}) requires that  
we determine the pairs of weights $p,p'$ that satisfy the following parity
equation
\beq
\e_n(\G;hp) \, \e_n(\G;hp') = \prod_{\alpha > 0}\, \e_n(\alpha \cdot hp) \,
\e_n(\alpha \cdot hp') = +1, \qquad \hbox{for all $h$ in $\Z^*_{nD}$}.
\label{affine}
\eeq
{}From the formula (\ref{sine}), this is equivalent to solve
\beq
\s_h\Big(\prod_{\alpha > 0}\, \sin{\pi\alpha \cdot p \over n}
\, \sin{\pi\alpha \cdot p' \over n}\Big) = \prod_{\alpha > 0} \,
\sin{\pi h \alpha \cdot p \over n} \, \sin{\pi h \alpha \cdot p' \over n}
> 0, \quad \forall h \in \Z^*_{nD}.
\eeq
In other words, the positive algebraic real number within the brackets in the
l.h.s. must be such that its Galois conjugates are all positive. Such numbers
are called totally positive. The previous equation can thus be interpreted by
saying that $p,p' \in P_{++}(\G)$ satisfy the parity rule
iff $S_{\rho,p}\,S_{\rho,p'}$ is totally positive.

Obviously, sums, products and ratios of totally positive numbers are totally
positive. A classical theorem about totally positive numbers is due to Landau
and Hilbert (see e.g. \cite{siegel}).
\begin{theo}
A real algebraic number $\xi$ is totally positive if and only
if it is a sum of squares in $\Q(\xi)$.
\end{theo}
The proof is easy. If $\xi$ is a sum of squares, it is immediate
that it is totally positive. Conversely, assume that $\xi$
is totally positive. Let $P(x)$ be the minimal polynomial
of $\xi$:
\beq
P(x)=x^n-a_1 x^{n-1}+a_2 x^{n-2}+\cdots + (-1)^n a_n.
\eeq
Then the rational numbers $a_i$ are all non-negative. The condition $P(\xi)=0$
can be written:
\beq
\xi(a_{n-1}+a_{n-3}\xi^2+\cdots) = a_n + a_{n-2}\xi^2+\cdots
\eeq
Set $\nu=a_{n-1}+a_{n-3}\xi^2+\cdots$. Observe that $\nu\neq 0$ by the
minimality of $P(x)$.
Then we have:
\beq
\xi=\frac{1}{\nu^2}(a_{n-1}+a_{n-3}\xi^2+\cdots)(a_n + a_{n-2}\xi^2+\cdots)=
\frac{1}{\nu^2}(b_0+b_1 \xi^2+\cdots),
\eeq
where the $b_i$ are positive rationals. Since a positive rational is easily
seen to be a sum of rational squares, the proof is complete. \cqfd

Thus in order to solve the parity equation for affine algebras, we look for
products of sines, in even number, which can be written as sums of squares in
$\Q(\sin{\pi \over n})$. 

For $n$ an integer and $d$ a divisor of $n$, the identity $1-X^d =
\prod_{j=0}^{d-1} \, (1-\zeta_d^jX)$ implies a number of product relations
labelled by an integer $a$
\beq
\sin{\pi ad \over n} \, \prod_{j=0}^{d-1} \, \sin{\pi(a + jn/d)
\over n} = 2^{1-d} \, (\sin{\pi ad \over n})^2, \qquad d|n,\; 1 \leq a \leq d-1.
\label{sines}
\eeq
The right--hand side is manisfestly totally positive, and so is the left--hand
side:
\beq
\s_h\Big(\sin{\pi ad \over n} \, \prod_{j=0}^{d-1} \, \sin{\pi(a + jn/d)
\over n}\Big) > 0.
\label{sigmasine}
\eeq
In order to convert this statement into identities involving parities, one
simply remembers that $\sin{\pi x \over n}$ lies in $\Bbb
Q(\zeta_{4n})$\footnote{Indeed, $\sin{\pi x \over n} = -{i \over
2}(\z_{2n}^x-\z_{2n}^{-x}) = -\half (\z_{4n}^{2x+n}-\z_{4n}^{-2x-n})$.}, so that
the Galois group acts on it by
\beq
\s_h(\sin{\pi x \over n}) = i\s_h(-i) \, \sin{\pi hx \over n}
= i\s_h(-i) \, \e_n(hx) \, \sin{\pi \restn{hx} \over n}.
\eeq
Thus the positivity of a Galois conjugate is not only determined by an $su(2)$
parity, but can be affected by a sign $i \s_h(-i)$. These signs (which depend
on $h$) drop out when $\s_h$ acts on an even number of sines, but otherwise
give extra contributions when the number of sines is odd.

If $d$ is odd, the number of sines is even, and (\ref{sigmasine}) leads to
identities between $su(2)$ parities 
\beq
R_n(d,a) \equiv \e_n(had) \, \prod_{j=0}^{d-1} \, \e_n(ha + hjn/d) = +1, \qquad
\forall h \in \Z^*_n\,, \;\hbox{$d$ odd}.
\label{dodd}
\eeq

If $d$ is even, we multiply the identity (\ref{sines}) by a positive rational
sine, say $\sin{\pi f \over n} \in \Q$, thereby preserving the total positivity.
The resulting identities now involve an even number of sines, and can be turned
into identities among parities 
\beq
R_n(d,a,f) \equiv \e_n(hf)\,\e_n(had) \, \prod_{j=0}^{d-1} \, \e_n(ha + hjn/d) =
+1, \qquad  \forall h \in \Z^*_n\,, \;\hbox{$d$ even}.
\label{deven}
\eeq
The allowed values $f={n\over 2},{n\over 6}$ and $5n \over 6$ are the
only rationals such that $\sin{\pi f \over n}$ is a strictly positive rational
number.

Thus we have succeeded in writing many identities $R_n(d,a)$ and
$R_n(d,a,f)$ involving $su(2)$ parities, which can be used to give solutions to
the parity equation in affine algebras. Here the main problem is precisely to
recast these identities in the form (\ref{affine}), in which the arguments of
the parities are related to the weights $p,p'$ in a very specific way. It is
nevertheless instructive to see how the known solutions of the parity equation
can be understood in terms of the above relations.

First of all, because the parity function for $\G$ is a product of parities for
$su(2)$, one can solve the parity equation (\ref{affine}) by equating the
$\e_n$'s by pairs. These rather trivial solutions can lead to non--trivial
couplings in terms of the weights, and it turns out that many apparently
non--trivial couplings are in fact trivial in this sense. For instance in
$su(5)$, it was found in \cite{rtw}, and checked the hard way, that the identity
$p=(1,1,1,1)$ can couple, for even $n$, to the following three weights
$p'=(1,{n \over 2}-2,{n \over 2}-2,1)$, $({n \over 2}-3,1,1,{n \over 2}-3)$ and
$({n \over 2}-3,2,2,{n \over 2}-3)$. To see that these three weights
indeed satisfy the parity equation with $p$ amounts to verify respectively the
identities 
\bea
&& \e_n(2h) \, \e_n(n-2h) \, \e_n(4h) \, \e_n(n-4h) = +1, \qquad \forall h, \\
&& \e_n(4h) \, \e_n(n-4h) = +1, \qquad \forall h, \\
&& \e_n(2h) \, \e_n(n-2h) = +1, \qquad \forall h,
\eea
simple consequences of the symmetry (\ref{auto}) of the function $\e_n$.
These three couplings appear in the $su(5)$ exceptional invariants due to 
conformal embeddings, at height $n=8,10$ and 12.

Many of the allowed couplings which are not trivial in the sense of the
previous paragraph follow from the relations (\ref{dodd}) and (\ref{deven}).
For instance in $su(3)$ at height $n$, the coupling of $(1,1)$ to $(1,{n
\over 2})$ is allowed due to the identity
\beq
\e_n(h) \, \e_n(2h) \, \e_n({\textstyle {nh \over 2}}) \, \e_n({\textstyle {nh
\over 2}}+h) = +1,
\eeq
which is nothing but the identity $R_n(2,1,{n\over 2})$. Similarly the
coupling of $(1,2)$ to $(2,{n\over 3}-1)$ is a consequence of
$R_n(3,1)$. Aoki \cite{aoki} has determined, for all integers $n$ except 32
values between 3 and 180, all pairs $p,p'$ of $su(3)$ weights which satisfy the
parity equation. His result shows that, besides the trivial solutions, all the
others follow from the identities (\ref{dodd}) and (\ref{deven}), and products
thereof. The same pattern holds in higher rank algebras, and points to the
genericity of the solutions provided by these identities. That they do
not exhaust the solutions follows from a concrete example: in $su(3)$ at height
$n=15$, the weights $(1,1)$ and $(1,5)$ are allowed to couple, due to the
identity
\beq
\e_{15}(h) \, \e_{15}(2h) \, \e_{15}(5h) \, \e_{15}(6h) = +1,
\eeq
which does not seem to follow from the product relations $R_n$.

The use of these to solve parity equations for affine algebras remains a
delicate matter, as subtle cancellations among individual parities must take
place. A good (but still mild) illustration of this is provided by
$su(4)$ at height $n=14$, where there is a coupling between $(1,1,1)$ and
$(1,2,7)$, due to three mechanisms: cancellations of pairs of identical $\e_n$,
the symmetry $\e_n(x)=\e_n(n-x)$ and the relation $R_{14}(2,2,7)$. 


\section{Bernoulli numbers}

In this section, we propose a second approach, based on the additive formulas
of Section 2. It is not entirely new, since the corresponding formula
(\ref{a2add}) for $su(3)$ was at the root of the works of Aoki \cite{aoki}, and
of Koblitz and Rohrlich \cite{kr}. With the additive formulas developed
in Section 2, the method can be extended to any affine algebra. The new feature
that appears when one goes beyond $su(2)$ and $su(3)$, is the presence of a
congruence (all expressions are valued in a finite ring). As we shall see,
this is the source of difficult arithmetical problems, which somehow embody the
difficulties inherent to high rank algebras. 

Our purpose here is not to report on the results we have obtained so far by
following this approach, since they are not conclusive at the moment. They
however suggest that this path is well suited for dealing with higher algebras.
Here we will briefly explain the method and give a feeling for the problems that
arise. A detailed and more complete account will appear elsewhere.

The parity equation, expressing the equality of a number of parities
$\e_n(\G;hp) = \e_n(\G;hp')$, is what we want to solve. The additive formulas,
like those of Prop. 1 in Section 2, give an expression for each of these
parities as a sum of residues modulo some integer. Thus the typical
problem is to find, for given and fixed $n$, all integers $x_i,y_i$ satisfying:
\beq
\sum_i \; \restn{hx_i} = \sum_i \; \restn{hy_i} \quad \bmod 2n, \qquad \forall
h \in Z^*_n.
\label{basic}
\eeq
The integers $x_i,y_i$ will eventually be related to the weights $p$ and $p'$
through their scalar products with positive roots of $\G$ (and so are not all
independent).

The basic idea is to write the equation (\ref{basic}) in the basis of
characters of $Z^*_n$, so we begin by recalling what these are. 

Characters modulo $n$ are homomorphisms of the multiplicative group $\Z^*_n$,
{\it i.e.} they are multiplicative functions $\t$, satisfying
$\t(hh')=\t(h)\t(h')$ for all $h,h' \in Z^*_n$, and of norm equal to 1. In 
concrete terms, if we write $\Z^*_n = \times_i \; Z_{m_i}$ as a product of
cyclic groups, every element can be uniquely expressed as $h=\prod_i
\;g_i^{t_i}$, with $g_i$ a generator of $Z_{m_i}$. An arbitrary character is
labelled by a set of integers $a_i$, taken modulo $m_i$, and takes the simple
form
\beq
\t(h) = \z_{m_1}^{a_1t_1} \, \z_{m_2}^{a_2t_2} \, \ldots, \qquad
0 \leq a_i \leq m_i-1.
\label{char}
\eeq
The character is even or odd depending on whether $\t(-1)=+1$ or $-1$. If
all $m_i$ are chosen to be even integers, a character being even or odd means
$\sum_i \; a_i = 0$ or 1 modulo 2.

A character of $Z^*_n$ may be extended to $Z_n$ (the set of all integers modulo
$n$), by setting $\t(t)=0$ if $t$ is not in $Z^*_n$. If $n \,|\, N$, it may be
further lifted to $Z_N$ by periodicity modulo $n$ (not forgetting the
coprimality condition\footnote{For instance, the character modulo 3 defined
by $\t(1)=1$, $\t(2)=-1$, can be extended modulo 6 by setting $\t(1)=1$,
$\t(5)=-1$.}), in which case we say that the resulting character of $Z_N$ is
induced by a character of $Z_n$. A character of $Z_n$ is called primitive if it
is not induced by a character of a subgroup of $Z_n$. A character modulo $n$ is
said to have conductor $f$ if it is induced by a primitive character modulo $f$
(so $f \,|\, n$). Loosely speaking, a character of conductor $f$ truncates its
argument modulo $f$, and so the conductor of a character is its period. 

Let us come back to the parity equation (\ref{basic}). It says that
\beq
\sum_i \; \restn{hx_i} - \sum_i \; \restn{hy_i} = 2n F(h\,|\,x_i,y_i),
\eeq
for some integral function $F$. Because $\restn{-x}=n-\restn{x}$, the
left--hand side is an odd function of $h$, and so is the function $F$.
Multiplying by $\t(h)$, a character modulo $n$, and summing over $h$ yields zero
if $\t$ is an even character, while it gives a multiple of 2 if $\t$ is
odd\footnote{By this is meant that $\sum_h \; F(h\,|\,x_i,y_i)\,\t(h)$ is an
algebraic integer, lying in the principal ideal (2) of some cyclotomic integer
ring.}. One obtains
\beq
\sum_i \; \sum_{h \in \Z^*_n} \; \restn{hx_i} \, \t(h) - 
\sum_i \; \sum_{h \in \Z^*_n} \; \restn{hy_i} \, \t(h) = 0 \quad \bmod 4n.
\label{cong}
\eeq
The change from a congruence modulo $2n$ to one modulo $4n$ is crucial for what
follows. 

It is important to realize that the equation (\ref{cong}) takes place in the
ring of integers of the cyclotomic extension $\Q(\z_{\varphi(n)})$ (containing
the values of $\t$). Thus the congruence involved is a condition
in the finite ring $\Z(\z_{\varphi(n)})/(4n)$. By a previous remark, it is
identically satisfied if $\t$ is an even character, so from now on, we
concentrate on the odd ones.

The equation (\ref{cong}) is a sum of terms of the form $\sum_h
\;\restn{hx}\,\t(h)$. Let us first compute this number when $x$ is coprime with 
$n$ (invertible modulo $n$). For convenience, we include a factor $1 \over n$,
and obtain, by a simple change of variable,
\beq
{1 \over n} \; \sum_{h \bmod n} \;\restn{hx}\,\t(h) = {1 \over n} \; 
\sum_{t \bmod n} \; \restn{t}\,\t(x^{-1}t) = \t^*(x) \, B_{1,\t}^n,
\eeq
where $B_{1,\t}^n$ is a generalized Bernoulli number (see f.i. \cite{wash})
\beq
B_{1,\t}^n = {1\over n} \; \sum_{t=1}^n \; t\,\t(t).
\label{bern}
\eeq
If $x$ is not coprime with $n$, the calculation is only slightly more
complicated. If we set GCD($x,n)={n \over e}$ and $\tilde x={x \over (n/e)}$
(so that $\tilde x$ is coprime with $e$), a little calculation shows that for a
character modulo $n$ of conductor $f$, the above sum is equal to
\beq
{1 \over n} \; \sum_{h \bmod n} \;\restn{hx}\,\t(h) = 
\cases{ 0 & if $f \nmid e$, \cr \noalign{\medskip}
{\varphi(n) \over \varphi(e)} \, B_{1,\t}^e \, \t^*(\tilde x) & if $f \mid e$.}
\label{fund}
\eeq

Using these results, the parity equation in the form (\ref{cong}) is a
congruence modulo 4 (we have divided by $n$) for a sum of terms comprising
Bernoulli numbers, various factors related to gcd's, and values of characters.
Instead of writing the complete equation in the general case, which does not
pose any problem but the notation, we take a simple example, and write it
explicitely in the case of $su(4)$. 

To simplify a bit more, we take in $su(4)$ two self--conjugate weights
$(a,b,a)$ and $(a',b',a')$, and assume that $a,b,2a+b,a',b',2a'+b'$ are all
coprime with $n$ (this last assumption simplifies the notation, but is actually
the most difficult situation). From (\ref{su4}), the congruences we must solve
are simple to write out
\beq
{\textstyle \half} B_{1,\t}^n\,\Big[2\t^*(a) + \t^*(b) - \t^*(2a+b) - 2\t^*(a')
- \t^*(b') + \t^*(2a'+b')\Big] = 0 \;\; \bmod 2, \quad \hbox{for all odd $\t$}.
\label{congsu4}
\eeq
Solving them requires looking more closely at the Bernoulli numbers. 

As it turns out, Bernoulli numbers have received an enormous attention for
decades, because of the extremely important role they play in algebraic number
theory. It would be an impossible task for us to make a review of their
properties. Instead, we will mention, without proof\footnote{For some of the
results mentioned in the text, we have provided our own proof, although we have
no doubt that they can be found somewhere in the mathematical literature.},
those which we feel are relevant for our problem. 

A first observation is that the congruence (\ref{congsu4}) is between algebraic
integers. The reason is very simple. The first congruence we wrote down,
equation (\ref{basic}), is the equality of two sums of residues, which are equal 
to 0 or to $n$ modulo $2n$ (as follows from the lemma of Section 2). But since
in any case, they are both equal to 0 modulo $n$, the congruence (\ref{basic})
is in fact trivial modulo $n$. When multiplied by $\t(h)$ and summed over $h$,
it yields (\ref{cong}), which must therefore be identically satisfied modulo
$2n$. It means that the equation (\ref{congsu4}) is identically
satisfied modulo 1, {\it i.e.} that the left--hand side is an algebraic
integer. Thus the non--trivial part is entirely contained in a congruence
modulo 2. 

Technically, these observations are reflected by specific properties of the
Bernoulli numbers $B_{1,\t}^n$. Indeed, one can show that most of them are not
only algebraic integers \cite{carlitz}, despite the factor $1 \over n$ in their
definition, but are also equal to 0 modulo 2. In other words, many numbers
$\half B_{1,\t}^n$ are integral. The precise conditions under which this is
true are not simple to state, but a sufficient condition is that the conductor
of $\t$ should not be a prime power\footnote{A particular instance where it is
not true is when $n$ is a power of an odd prime $p$. Then $B_{1,\t}^n$ is not 
integral, but there is a unique prime ideal $\pi$ in $\Q(\z_{\varphi(f)})$,
lying above $p$, such that $\pi B_{1,\t}^n$ is integral. In this situation,
the announced triviality of the congruence modulo $n$ is fulfilled because the
various characters in (\ref{congsu4}) add up to something equal to 0 modulo
$\pi$.}. 

When $\t$ is such that $\half B_{1,\t}^n$ is integral, the equation
(\ref{congsu4}) simplifies further to become
\beq
{\textstyle \half} B_{1,\t}^n \, \Big[\t^*(b) + \t^*(2a+b) + \t^*(b') +
\t^*(2a'+b')\Big] = 0 \;\; \bmod 2.
\label{congsu4_2}
\eeq

The main difficulty that arises when one tries to solve equations like
(\ref{congsu4}) or the previous one, is to calculate the gcd of ${\textstyle
\half} B_{1,\t}^n$ and 2. Clearly the most favourable case is when the two
numbers are coprime, because one can then divide by ${\textstyle \half}
B_{1,\t}^n$ and study the conditions under which the sum of characters
vanishes. Although that part may not be straightforward, we think it should be
tractable, since after all, it is merely a matter of having a certain sum of
roots of unity that vanishes. Even if exotic solutions can occur, the
generic solutions are expected to be the trivial ones, namely $a'=a$ and $b'=b$
(up to some automorphisms). 

To see if half the Bernoulli numbers are coprime with 2, and if not, to
calculate their gcd, is much more delicate. Even worse is the fact that they
can vanish (as complex numbers). Indeed a standard identity gives the 
Bernoulli numbers associated to non--primitive characters in terms of those
pertaining to primitive characters. If $\t$ has conductor $f$, and if
$\t_0$ is the character modulo $f$ that induces $\t$, then the formula is
\cite{wash}
\beq
B_{1,\t}^n = B_{1,\t_0}^f \; \prod_{{\rm prime} \; p \mid n} \, (1 - \t_0(p)).
\eeq
It is known that Bernoulli numbers associated to primitive characters are
non--zero as complex numbers, so $B_{1,\t_0}^f \neq 0$, but the product over
the prime divisors of $n$ may force a zero (this can only happen if $n$ is not
a prime power). As to the congruence modulo 2, ${\textstyle \half} B_{1,\t}^n$
can have a common divisor with 2, either because ${\textstyle \half}
B_{1,\t_0}^f$ has one, or else because some $(1 - \t_0(p))$ divides 2. All these
questions lead to rather non--trivial arithmetical questions in cyclotomic
extensions.

It is however intriguing to note that the generalized Bernoulli numbers appear
in a remarkable formula expressing what is called the relative class number
$h^-$ of cyclotomic fields. If $h_n$ and $h^+_n$ denote respectively the class
number\footnote{If $\Bbb K$ is a number field, {\it i.e.} a finite algebraic
extension of $\Q$, the fractional ideals of $\Bbb K$ form an Abelian group,
where the identity is just the ring of integers of $\Bbb K$. One defines an
equivalence relation by saying that two ideals $\alpha$ and
$\beta$ are equivalent if $\alpha \beta^{-1}$ is principal (generated
by a single element of $\Bbb K$). One can show that the quotient of the group of
ideals by this relation is a finite group, called the ideal class group. Its
order is the class number of $\Bbb K$, and is among the most important numbers
characterizing $\Bbb K$.} of $\Q(\z_n)$ and of $\Q(\cos{2\pi \over n})$, the
relative class number of $\Q(\z_n)$ is their quotient, $h^-_n = h_n/h^+_n$. This
number, also an integer as it turns out, can be computed from the formula
\cite{wash}
\beq
h^-_n = \tilde Q n  \prod_{\t \; {\rm odd} \atop {\rm  primitive}} 
(-{\textstyle \half} B_{1,\t}^n),
\eeq
where $\tilde Q$ is a numerical factor, equal to 1 if $n$ is a power of 2, 2 if
$n$ is a odd prime power or if $n$ is even, and 4 otherwise. From this formula,
one can see that to determine the GCD of $\half B_{1,\t}^n$ and 2 amounts to say
something about the power of 2 that divides the relative class number of
cyclotomic fields. In this respect, Iwasawa's theory of $\Z_p$--extensions
could provide some help.

Certainly, one cannot hide the fact that hard and maybe deep problems
lie on the way towards the solution of the parity equation. However, one should
emphasize the fact that these problems, mostly concerned with Bernoulli
numbers, are not specific to the $su(4)$ situation that we chose as
illustration. If one follows the approach presented here, be it in $su(4)$ or
in another algebra, one ends up with equations like (\ref{congsu4}) or
(\ref{congsu4_2}), the solution of which requires basically two steps.
One involves the Bernoulli numbers themselves, more precisely their
modular properties; the other is an equation saying
that certain values of characters add up to zero. Only this second part depends
on which algebra we treat and which kind of weights. The first part is
universal, algebra independent. This may be a happy coincidence as it is
probably more difficult. 

We can illustrate this by displaying the analogous
equation\footnote{Interestingly, if we take two  self--conjugate weights of
$su(5)$, we obtain the same equation as for $su(4)$ (with $b$ replaced by
$2b$): the two weights $(a,2b,a)$ and $(a',2b',a')$ satisfy the parity equation
for $su(4)$ if $(a,b,b,a)$ and
$(a',b',b',a')$ satisfy the parity equation for $su(5)$. One easily convinces
oneself that the same holds within all pairs of algebras $su(4\ell)$ and
$su(4\ell+1)$, if one restricts to self-conjugate weights.} for
$su(8)$, at height $n$. We make the same assumptions as for $su(4)$, namely we
take two self--conjugate weights $(a,b,c,d,c,b,a)$ and $(a',b',c',d',c',b',a')$.
As before we assume that all linear combinations of the Dynkin labels that
appear are coprime with $n$. Then the equivalent of (\ref{congsu4_2}) involves
a sum of only eight characters
\beq
{\textstyle \half} B_{1,\t}^n \, \Big[\t^*(d) + \t^*(2c+d) + \t^*(2b+2c+d)
+ \t^*(2a+2b+2c+d) + \hbox{same primed} \Big] = 0 \;\;\bmod 2, \qquad 
\eeq
valid for all odd characters which are such that ${\textstyle \half}
B_{1,\t}^n$ is integral.  

Without minimizing the difficulties, we believe that it is a very positive and
encouraging feature of the approach presented here.


\section*{Acknowledgements}

D.A. would like to thank the Institute for Theoretical Physics in
Louvain--la--Neuve for hospitality extended to him during visits. This work was
completed while he was staying at the Institute of Theoretical Physics, ETH
Z\"urich, and was supported by the Fonds National Suisse. He thanks both
institutions.


\end{document}